\documentclass[times]{dacauth}

\usepackage{moreverb}

\usepackage[dvips,colorlinks,bookmarksopen,bookmarksnumbered,citecolor=red,urlcolor=red]{hyperref}
\usepackage{algorithm}
\usepackage{algorithmic}
\usepackage{amsfonts,amsmath,latexsym,amssymb}
\usepackage{graphicx}
\usepackage{CJK}
\usepackage{multirow}
\usepackage{chemarrow}
\usepackage{setspace}
\usepackage{epstopdf}
\usepackage{flushend}
\usepackage{amsthm}
\DeclareMathOperator*{\argmin}{argmin}

\newcommand\BibTeX{{\rmfamily B\kern-.05em \textsc{i\kern-.025em b}\kern-.08em
T\kern-.1667em\lower.7ex\hbox{E}\kern-.125emX}}

\def\volumeyear{2010}

\begin{document}


\title{Improving Reliability Performance of Diffusion-based Molecular Communication With Adaptive Threshold Variation Algorithm}

\author{Peng He\affil{1}, Yuming Mao\affil{1}, Qiang Liu\affil{1} and Kun Yang\affil{2}}

\address{\affilnum{1}School of Communication and Information Engineering, University of Electronic Science and Technology of China, Chengdu, China\break
\affilnum{2}Network Convergence Laboratory, School of Computer Science and Electronic Engineering, University of Essex, Colchester, UK}

\corraddr{Peng He, School of Communication and Information Engineering, University of Electronic Science and Technology of China, Chengdu, China\\
\corrauth E-mail: hp6500@126.com}

\begin{abstract}
In this work, we investigate the communication reliability for diffusion-based molecular communication, using the indicator of bit error rate (BER). A molecular classified model is established to divide molecules into three parts, which are the signal, inter-symbol interference (ISI) and noise. We expand each part separately using molecular absorbing probability, and connect them by a traditional-like formula. Based on the classified model, we do a theoretical analysis to prove the feasibility of improving the BER performance. Accordingly, an adaptive threshold variation (ATV) algorithm is designed in demodulation to implement the goal, which makes the receiver adapt the channel condition properly through learning process. Moreover, the complexity of ATV is calculated and its performance in various noisy channel is discussed. An expression of Signal to Interference plus Noise Ratio (SINR) is defined to verify the system performance. We test some important parameters of the channel model, as well as the ATV algorithm in the simulation section. The results have shown the performance gain of the proposal.
\end{abstract}


\maketitle


\vspace{-6pt}

\section{INTRODUCTION}
Molecular communication (MC) is an attractive domain which uses the molecules as the carriers to permit the information exchange between nano-devices at nanoscale [1]. MC has the advantage of energy-saving and no-radiation, which is more suited in biological environment compared with the traditional communication. Rapid growth of nano-technology promotes the manufacture of bio-inspired nano-devices, which are able to perform basic tasks, including sensing, actuating and computing. Limited by size and power, multiple nano-devices are essential to work together as a network to execute a complex task, composing the internet of nano-things. One typical case is the body area network, in which the nano-devices enjoy a bright prospect on nano-healthcare [2], and will gradually take place of traditional applications, such as the wireless sensor network [3,4]. Some another potential application domains include environmental monitoring, industry and military equipment, respectively described in [5]-[7].

There are some significant differences between traditional communication and MC, such as the information carrier, transmission speed, range, noise source and so on. MC could also be divided into various forms according to specific rules. Comparing with other forms in MC, diffusion-based MC is inspired by drift flows or relies on molecular thermodynamics movement, which derives from the natural cases and is supported by fluid and molecular mechanics. Diffusion-based MC is worthy to be investigated not only in communication, because it could be applied in many useful applications. One typical example is the drug delivery [8,9], in which particles are released from the sickness, diffused in the blood medium for accurate positioning. Hence, we adopt diffusion-based MC as the research scenario in our paper.

Communication reliability is an important issue all the time. Considering the randomness and uncertainty of molecular movement, MC suffers a more serious problem compared with traditional communication. At present, bit error rate (BER) is chosen as the indicator of the reliability in MC generally. For example, [10] does a brief investigation of reliability using BER in diffusion-based MC, but lacks deep analysis as the specific channel characteristics. [11] proposes the forward error correction codes over MC and verifies by BER. [12] designs the receiver with different signal detection technologies, in which the performances of those proposals are shown using BER.

In our work, on-off key (OOK) modulation is adopted as most works do. Assuming transceivers are strictly time synchronous, one time slot is utilized to transmit one bit. We first introduce the basic system model and communication process. Then, based on the molecular absorbing probability, we propose a classified model to divide the molecules into signal, inter-symbol interference (ISI) and noise branches. Inspired by traditional communications, we establish a brief formula to connect the three branches. We expand each branches in detail, and give the expression of BER.

In our work, to improve the performance of BER, we propose an algorithm named adaptive threshold variation (ATV). A relevant work is introduced in [13], where an adaptive transmission rate method is designed by fixing the receiver threshold and altering the transmission rate of transmitter. Inversely, our ATV algorithm fixes the transmission rate and alters the receiver threshold adaptively. We first prove the feasibility of the ATV design through 3 Propositions in theoretical analysis. Then we design ATV in detail. In ATV, the threshold varies based on the knowledge of the previous bits, which is related on the channel condition. The threshold variation is similar with the learning process, that make the receiver adapt the channel for the demodulation. We calculate the complexity of the ATV and discuss its operation under various noisy conditions. In the simulation section, we deduce the expression of signal to interference plus noise ratio (SINR). The results show that the optimal threshold is not at the general midpoint of the molecule quantity per bit, which is consistent with the theoretical analysis. ATV algorithm is verified in the SINR-BER curve, indicating that there is a significant performance improvement. In addition, we show the threshold variation for various algorithm parameter and under different communication conditions, to give the guidance of algorithm setting.

The remainder of this paper is organized as follows. In section 2, we list the related work in this field. In section 3, we introduce the basic system model of the diffusion-based MC. Then, the molecular classified model is proposed in section 4. In this section, signal, ISI and noise are respectively analyzed. In section 5, the ATV algorithm is analyzed, designed and discussed in detail. The numerical results are presented in section 6, where we evaluate the BER performance in the analytical model, as well as the algorithm performance. We conclude this paper in section 7.

\vspace{-6pt}

\section{RELATED WORK}
\vspace{-2pt}

Molecular communication is promising to be utilized in body environment, implemented based on the biological channels. According to different mediums, MC proceeds with different forms. Some potential forms under study includes diffusion-based MC [14], neural communication [15], molecule motor MC [16], blood vessel communication [17], bacteria-based communication [18], etc. Among those forms, diffusion-based MC is a typical one, which is similar with the traditional wireless communication. Major efforts in the area of diffusion-based MC focus on the study of physical layer at present.

Diffusion-based molecular communication is usually a composite physical process, driven by two different motivations. For the former, molecules in the environment perform the Brownian movement [14]. For the latter, molecules are driven by active drift flow [19]. Most of the works adopt synchronous mechanism, i.e., time slots are used to transmit bits. Differently, [20] proposes an asynchronous method for MC. For channel estimation, some works focus on the molecular concentration in the environment, which follows the well-known Fick's second law [1]. Another works investigate the random walk of solo molecule, and model the distribution by winner process [19]. To evaluate the communication performance, some important indicators of diffusion-based MC system, like channel capacity [21], delay [22], bit error rate [10], are exploited based on the knowledge of traditional communication and information theory.

Diffusion-based MC suffers serious interference problem in time-slotted system because of the randomness of molecule moving, even the regularities of distribution could be tracked. Inter-symbol interference (ISI), one of the typical interference in MC, are caused if molecules of other bits impact the current bit [23]. Some methods to eliminate ISI are designed accordingly. For instance, [24] tries to use the enzyme to remove the redundant molecules, and alleviate the interference between different slots. A decision back decoding method is proposed in [25], which could eliminate ISI based on the calculation of previous bits. The noise of MC is determined by several factors. [26] does a deep analysis of noise for diffusion-based MC. The choice of carrier molecules is a major one, because there may exist various background molecules in the environment, such as the human body. A specific molecular type should be chosen as the information carrier to alleviate the effect of those environmental molecules. Another main factor is interference from another nano-devices, for a nanoscale device to device (D2D) communication, this could be classified into noise source in MC.

Many modulation methods of MC are extended from the traditional ones, including AM [27], FM [28], PM modulation [20], etc. In MC, AM modulation is based on the concentration of molecules, which maps potential of traditional communication. "01" bits are obtained via comparison between concentration and a decision threshold. Among AM modulations, OOK is the simplest way and adopted in most MC literatures. FM modulation utilizes different types of molecules, which map various frequency in traditional communication. The interference between types of molecules comes from the possible chemical reactions, so it's important to choose the appropriate types. PM modulation of MC is based on the release time of molecules. In one time slot, the molecules could be released at different time, that lead the difference of the phase.

\begin{figure}
\centering
\includegraphics[width=3.5in]{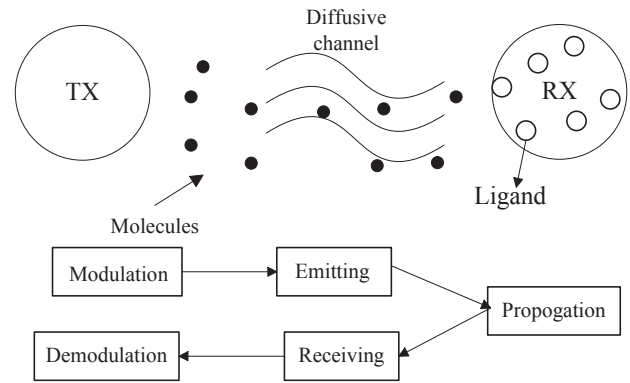}
\caption{The diffusion-based MC system}
\end{figure}

\section{SYSTEM MODEL}
In this paper, one transmitter (TX) and one receiver (RX) nano-devices are considered, with the fixed distance denoted by $r$, as Figure 1 shown. The essential communication process contains modulation, emitting, propagation, receiving and demodulation. We assume that the transceiver is strictly time synchronous. A major goal of MC is to realize the reliable transmission of binary strings, which is denoted by $b(i)$ in our model. Each bit is emitted in a transmitting slot $n_t$ and is respected to be absorbed in a corresponding receiving slot $n_r$. Those time slots have the fixed length, given by $\tau$. One example diagram of transceiver slots mapping is shown in Figure 2.

\begin{figure}
\centering
\includegraphics[width=3.5in]{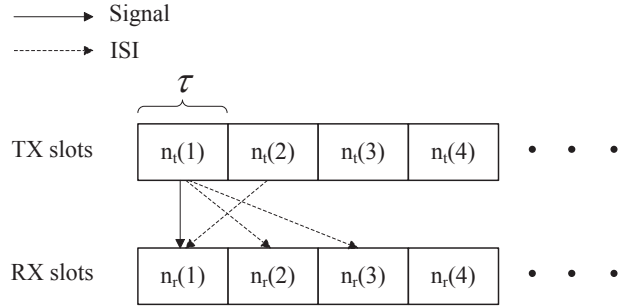}
\caption{Transceiver slots mapping diagram}
\end{figure}

\subsection{Modulation and emitting}

Modulation is the first process of the communication. On one hand, restrained by the tiny size and limited power of nano-devices, it's not easy to achieve some complex communication technologies. One the other hand, many natural cells usually release specific molecules when communication is necessary, and not release molecules during the rest time. So the on-off key (OOK) modulation could fit many natural cases. We express the OOK modulation as,

\begin{equation}
N_{tx}(i)=
\begin{cases}
M& \text{b(i) = "1"}\\
0& \text{b(i) = "0"}
\end{cases}
\end{equation}
Where $N_{tx}(i)$ is the released molecule quantity at the beginning of the $i^{th}$ slot. $M$ molecules are emitted for bit "1", no molecules are emitted for bit "0". Note that molecule quantity could not be a negative value, so it is a unipolar modulation.

\subsection{Propagation}
The movement of the molecules in the environment contains two folds. One is driven by active flows, the other is the random brownian movement, of which the composite mobility could be modeled by winner process. From the macroscopical view, molecular concentration is a useful indicator in MC, similar with the potential in traditional communication. Molecular concentration depends on several factors, including time, spatial position, transmission rate, channel condition, etc. In one-dimensional space, the molecular concentration distribution follows the Fick's second law [1], expressed as,

\begin{equation}
\frac{\partial c(x,t)}{\partial t}=D\Delta c(x,t)+r(0,t)
\end{equation}
Where $c(x,t)$ is the molecule concentration at location $x$ and time $t$. $\Delta c(x,t)$ means the sum of the second derivatives of $c(x,t)$. $r(0,t)$ is the transmission rate of the molecules at transmitter, of which the discrete form is given by $r(0,i\tau)=N_{tx}(i)$. $D$ is the diffusion coefficient related to some environment factors, given by,

\begin{equation}
D = \frac{k_B T}{6 \pi \eta R_H}
\end{equation}
Where $k_B$ is the Boltzman constant, $T$ is the environment temperature. $\eta$ is the dynamic viscosity of the fluid, and $R_H$ is the hydraulic radius of the molecules.

For various $r(0,t)$, different solutions are available for (2). The OOK modulation is adopted as above said, the solution of (2) could be the integral of molecular transmission rate and Green function $g(x,t)$ [29], which is a function of time and position, i.e.,

\begin{equation}
c(x,t)=r(0,t)*g(x,t)
\end{equation}

Note that $g(x,t)$ could also be regarded as the position distribution of a single molecule. In one-dimensional space, it is given by,
\begin{equation}
g(x,t)=\frac{1}{(4 \pi D t)^{3/2}} exp(-\frac{x^2}{4Dt})
\end{equation}

\subsection{Receiving and demodulation}

Let $V_R$ be the maximum absorbing space of the receiver. Apart from the noise, we calculate the molecules in this space during $i^{th}$ slot as,

\begin{equation}
N_{rx}(i)= \int_{i\tau}^{(i+1)\tau} c(r,t)*V_R dt
\end{equation}

In our model, we consider a realistic species of biological receiver, named ligand-based receiver. The specific ligand receptors are distributed uniformly on the surface of the receiver. It works in case that the carrier molecules are matched with the receptors. In other word, the receptors and signal molecules shall be chosen particularly, to be matched biologically. There are many cases in nature. For example, in neuron network, the receiver neural cells use specific ligand-based chemoreceptors to receive neurotransmitter in gap junction [30].
The receptors bind and release molecules continually, and (6) is rewritten as,

\begin{equation}
N_{rx}(i)=\int_{i\tau}^{(i+1)\tau} \frac{c(r,t)*V_R*a*Q}{b} dt
\end{equation}
Where $a$ and $b$ are the binding and releasing rate of receptors. $Q$ is the density of receptors on the surface of the receiver.

The receiver demodulates the information according to the absorbed molecule quantity $N_{rx}(i)$ of (7). Like traditional AM demodulation, a threshold is needed, that is fixed in general. We denote it by $N_T$. Bit "1" is obtained if $N_{rx}(i)$ is greater than threshold, else bit "0" is obtained. Hence, we get the error probability of $i^{th}$ bit for bit "1" as,
\begin{equation}
p_e^1(i)=\sum_{k=0}^{N_T-1}p(N_{rx}(i)=k)
\end{equation}

Similarly, the error probability of $i^{th}$ bit for bit "0" is,

\begin{equation}
p_e^0(i)=\sum_{k=N_T}^{\infty}p(N_{rx}(i)=k)
\end{equation}

In (8) and (9), $p_e^1(i)$ and $p_e^0(i)$ respectively means the error probability transmitting of "1" and "0". They correspond with the sum of probability which the estimation is "0" and "1". The expression of BER is,

\begin{equation}
p_e = p(b(i)=0) p_e^0(i) + p(b(i)=1) p_e^1(i)
\end{equation}
Where $p(b(i)=0)$ and $p(b(i)=1)$ are the probability of transmitting "0" and "1". In many cases, we assume that they are equal as 0.5.

\section{MOLECULAR CLASSIFIED MODEL}
In this section, we first give the absorbing probability of single molecule. Then, a classified model of molecules for demodulation is proposed, to divide the received molecules into signal, inter-symbol interference and noise parts. We establish a traditional-like formula to connect them, and expand them in detail.

\subsection{Molecular absorbing probability}
Considering that the distance between the transceiver is short (usually several $\mu m$), we assume that the molecules live long enough in the environment before absorbed by the receiver. The distribution of a single molecule is given by (5). We calculate its accumulative density function $G(x,t)$ as,

\begin{equation}
\begin{split}
G(x,t) &= \int_0^\infty g(x,t)dt\\
&= erfc(\sqrt{\frac{x^2}{4Dt}})
\end{split}
\end{equation}

Based on the theory of [32], the probability that a molecule is released in $i^{th}$ time slot and absorbed in $j^{th}$ time slot is,

\begin{equation}
\begin{split}
p'(i,j)&=\int_{(j-i)\tau}^{(j-i+1)\tau} g(r,t) dt\\
&=G(r,(j-i+1)\tau)-G(r,(j-i)\tau)
\end{split}
\end{equation}
Considering the ligand-based model, (12) can be revised as,

\begin{equation}
p(i,j)=\frac{a Q p'(i,j)}{b}
\end{equation}

\subsection{Molecular classification}
With the probability given above, we could calculate the molecule quantity that are absorbed by the receiver in various slots. According to this character, the molecules absorbed during one slot could be divided into three portions, i.e., signal molecules, inter-symbol interference (ISI) molecules and noise molecules. Similar with traditional communication, we establish a relation formula for the three portions,

\begin{equation}
p\{N_{rx}(n_r)=k\}=p\{N_{sig}(n_t,n_r)+\sum_{l=1,l \neq n_t}^\infty N_{isi}(l,n_r)+N_{noise}(n_r)=k\}
\end{equation}

In (14), the three portions of the right side respectively mean the expectative signal molecules, ISI molecules and noise molecules that are received during $n_r^{th}$ slot. We assume the corresponding quantities are respectively $k_0$, $k_1$ and $k_2$, s.t., $\sum_{i=0}^2 k_i=k$.

\subsubsection{Signal molecules}

The first term of (14) $N_{sig}(n_t,n_r)$ indicates the signal molecules in the $n_r^{th}$ slot, which are transmitted in $n_t^{th}$ slot. The relation between $n_t$ and $n_r$ mainly depends on channel condition, transmission distance and slot length. While every solo molecule in the environment performs a random movement, the macroscopical distribution could be calculated, which is expressed by molecular absorbing probability (13). The quantity of the signal molecules follows the binomial distribution [10],

\begin{equation}
p(N_{sig}(n_t,n_r)=k_0)=\left(\begin{array}{c} N_{tx}(n_t)\\k_0 \end{array}\right)p(n_t,n_r)^{k_0}(1-p(n_t,n_r))^{N_{tx}(n_t)-k_0}
\end{equation}

Notes that (15) is the signal molecules of bit "1". If $n_t^{th}$ bit is "0", (15) becomes 0.

\subsubsection{ISI molecules}

The inter-symbol interference (ISI) exists in the diffusive model also because of the randomness of molecular movement. As described before, each receiving slot maps a transmitting slot. We divide ISI into two themes, i.e., the interference from other bits and the interference to other bits. For a specific bit,
\begin{itemize}
\item The first theme indicates that the molecules of the non-transmitting slots are absorbed by receiver in receiving slot, they are considered as the ISI from other bits.
\item The second theme indicates that the molecules of the transmitting slot are absorbed by receiver in non-receiving slots, they are considered as the ISI to other bits.
\end{itemize}

The first theme increases the error rate of bit "0" demodulation. Because some molecules of the other bits are added in the demodulation of bit "0", increase the probability of exceeding the threshold to decode it to "1". In a similar way, we could get that the second theme increases the error rate of bit "1" demodulation.

An example could be listed in Figure 2, the dash line from $n_t(2)$ to $n_r(1)$ is the first theme, indicates the ISI from second bit to first bit. The dash line from $n_t(1)$ to $n_r(2)$ is the second theme, indicates the ISI from first bit to second bit. On condition that the length of slots $\tau$ is long, ISI of the adjacent bits is much heavier than that of the nonadjacent bits. So we only consider the ISI of the adjacent bits, which is presented by expanding the second item of (14),

\begin{equation}
\begin{split}
&p(N(n_t-1,n_r)+N(n_t+1,n_r)=k_1)\\
&=\sum_{h=0}^{k_1}p(N(n_t-1,n_r)=h)p(N(n_t+1,n_r)=k_1-h)
\end{split}
\end{equation}
Here, those of $(n_t-1)^{th}$ and $(n_t+1)^{th}$ are regarded as the interference molecules as a view of $n_r^{th}$ time slot. To further expanded, the expression is,

\begin{equation}
\begin{split}
p&(N(n_t-1,n_r)+N(n_t+1,n_r)=k_1)\\
=&\sum_{h=0}^{k_1} \left(\begin{array}{c} N_{tx}(n_t-1)\\h \end{array}\right)p(n_t-1,n_r)^h(1-p(n_t-1,n_r))^{N_{tx}(n_t-1)-h}\\
\times& \left(\begin{array}{c} N_{tx}(n_t+1)\\{k_1-h} \end{array}\right)p(n_t+1,n_r)^{k_1-h}(1-p(n_t+1,n_r))^{N_{tx}(n_t+1)-(k_1-h)}\\
\end{split}
\end{equation}

Note that ISI of $(n_t+1)^{th}$ bit exists only when $n_t\leq n_r$. The relation between $n_t$ and $n_r$ is determined by the transceiver. $n_t=n_r$ means that signal molecules which are emitted in one slot will be received in the same slot with a large probability, requiring that the distance between the transceiver is short or the slot length is large.

\subsubsection{Noise molecules}

The third item of (14) is the channel noise of the system. It is mainly caused by other nano-devices if nano-devices communication exist in the environment [28], apart from the possible molecular reaction and background molecules. When receiving and counting undesired molecules in demodulation, those molecules could be regarded as noise. In our literature, we cite the counting noise at the receiver proposed in [26]. The counting noise is accumulatively relevant to the quantity of nano-devices in the environment. More nano-devices will lead the heavier noise. On the contrary, if only one transceiver pair exists, the channel condition is much better. The direct impact of noise to the receiver is altering the quantity of the receiving molecules, i.e., the $N_{rx}(i)$ in (7). Positive noise means receiving redundant molecules from other nano-devices. Negative noise means some signal molecules are received by other nano-devices. We assume that the noise is Additive White Gaussian Noise (AWGN), which is similar with [28] and expressed as,

\begin{equation}
N_{noise}(n_r) \sim Normal(0,\sigma^2)
\end{equation}

In (18), $\sigma^2$ is the variance and indicate the noise power. It is positively related to the quantity of nano-devices and some other factors. Taking the expression of Normal distribution into account and the probability of third item in (14) is,

\begin{equation}
p(N_{noise}(n_r)=k_2)=\frac{1}{\sqrt{2 \pi}\sigma}e^{-\frac{(k_2)^2}{2\sigma^2}}
\end{equation}

\section{ATV ALGORITHM}
In this section, we first prove the feasibility of ATV algorithm in theory analysis, i.e., why we design ATV, based on the classified model of section 4. Then, we design the detail ATV algorithm based on the theory analysis, describe how it works. After that, we calculate the complexity of ATV, and discuss its performance on various channel conditions.

\subsection{Theory analysis}
The goal of the theory analysis is to prove the feasibility of ATV design. Theorem 1 below gives the theoretical optimal receiver threshold $N_T^{opt}$ in the demodulation of diffusion-based MC. Considering the impact of the random noise, the $N_T^{opt}$ changes with the channel condition, we denote its mean as $\overline{N_T^{opt}}$.

\newtheorem{therory1}{\textbf{Proposition}}

\begin{therory1}
Let $E\{N_{rx}^1\}$ and $E\{N_{rx}^0\}$ be the mean value of received molecules for bit "1" and "0", then $\overline{N_T^{opt}} = \frac{E\{N_{rx}^1\} + E\{N_{rx}^0\}}{2}$.
\end{therory1}

\begin{proof}
From (14) we could see that received molecules include signal, ISI and noise branches in the demodulation. Focusing on the noise term, $N_{noise}$ is a random Gaussian variable with zero-mean and variance of $\sigma^2$. Hence, for bit "1", we express the probability of absorbing $N_{rx}$ molecules as,
\begin{equation}
p(N_{rx}|1)=\frac{1}{\sqrt{2 \pi}\sigma}e^{-\frac{(N_{rx}^1-N_{isi}^1-N_{sig}^1)^2}{2\sigma^2}}
\end{equation}
In a similar way, for bit "0", we have,
\begin{equation}
p(N_{rx}|0)=\frac{1}{\sqrt{2 \pi}\sigma}e^{-\frac{(N_{rx}^0-N_{isi}^0-N_{sig}^0)^2}{2\sigma^2}}
\end{equation}

$N_T^{opt}$ should be the one that minimize BER performance in the demodulation, expressed by,

\begin{equation}
N_T^{opt}=\argmin \limits_{N_T \in [0,M]}p_e
\end{equation}

So $N_T = N_T^{opt}$ requires that $\frac{\partial p_e}{\partial N_0^{opt}}=0$, substituting (8) to (10), we have,

\begin{equation}
p(b(i)=1)p(N_T^{opt}|1)-p(b(i)=0)p(N_T^{opt}|0)=0
\end{equation}

Substituting (20) and (21), we calculate $N_T^{opt}$ as,

\begin{equation}
N_T^{opt} = \frac{N_{isi}^0+N_{sig}^0+N_{isi}^1+N_{sig}^1}{2}
+\frac{\sigma^2}{N_{isi}^1+N_{sig}^1-N_{isi}^0-N_{sig}^0}ln\frac{p(b(i)=0)}{p(b(i)=1)}
\end{equation}

In typical case, the probabilities of emitting "1" and "0" are equal.
And the mean value of $N_{noise}$ is 0, so we get its mean form as,

\begin{equation}
\overline{N_T^{opt}} = \frac{E\{N_{isi}^0\}+E\{N_{sig}^0\}+E\{N_{isi}^1\}+E\{N_{sig}^1\}}{2} = \frac{E\{N_{rx}^1\} + E\{N_{rx}^0\}}{2}
\end{equation}
\end{proof}

Note that $N_{rx}^1$ and $N_{rx}^0$ are both variables that are determined by multiple factors, such as the channel noise, medium type, slot length, etc. So the $N_T^{opt}$ is also a variable, that's why we calculate its mean. Slot length is an important factor in the system, which is related with the ISI and delay. In this paper, considering its impact on ISI,
we list the proposition 2 as follows,

\begin{therory1}
If $n_t=n_r$, $\tau > \frac{r^2}{6D}$ is a needed not sufficient condition of $N_{sig}>N_{isi}$.
\end{therory1}

\begin{proof}
According to the definition of signal and ISI molecules in section 4, they both follow the binomial distribution. When $n_t=n_r$, only the ISI from previous $n_t$ slots to later $n_r$ slots exists. So we get the mean value of $N_{isi}$ as $M\{G(2\tau)-G(\tau)\}$, based on the character of the binomial distribution. Similarly the mean value of $N_{sig}$ is $MG(\tau)$. So, we have,

\begin{equation}
N_{sig}>N_{isi} \rightarrow 2G(\tau)>G(2\tau)
\end{equation}

$G(\tau)$ and $G(2\tau)$ could be got for specific $\tau$ because it's the error function or Gauss error function, listed in (11). It's easy to calculate that to satisfy (26), $\tau > \frac{r^2}{6D}$ should be required.
\end{proof}

We choose $\frac{r^2}{6D}$ applied in theorem 2 because it's the time to peak of $g(x,t)$, that is important in demodulation and slot length designing. In theorem 3, we will prove the feasibility of ATV algorithm, with the acid of proposition 1 and 2.

\begin{therory1}
$\overline{N_T^{opt}}<\frac{M}{2}$, on condition that $n_t=n_r$.
\end{therory1}

\begin{proof}
Based on proposition 1, we get the expression of $\overline{N_0^{opt}}$ as,

\begin{equation}
\overline{N_T^{opt}} = \frac{E\{N_{sig}^1+N_{isi}^1+N_{sig}^0+N_{isi}^0+2N_{noise}\}}{2}
\end{equation}
Note that $E\{N_{noise}\}=0$, $E\{N_{sig}^0\}=0$, calculating the distribution of the remaining three items based on the binomial distribution, we have,
\begin{equation}
\overline{N_T^{opt}} = \frac{M\{2G(2\tau)-G(\tau)\}}{2}
\end{equation}
According to proposition 2, to make sure $N_{sig}>N_{isi}$, $\tau > \frac{r^2}{6D}$ is required. Under this condition , it's easy to know that $\overline{N_T^{opt}}$ of (28) is smaller than $\frac{M}{2}$.

\end{proof}

\subsection{ATV algorithm design}
At the receiver, the threshold $N_T$ is fixed generally for the demodulation of bit "1" or "0". As section 3 said, for typical demodulation, the molecules of one bit is received, if the quantity of receiving molecules $N_{rx}$ is greater than $N_T$, the estimation is "1". Conversely, the estimation is "0".

\begin{algorithm}[h]
\caption{Adaptive-threshold Variation}
\label{alg:Framwork}
\begin{algorithmic}[1]
\STATE Set $N_T(1) = M/2$, $n_1=0$, $n_0=0$, $N_{rx}^1=0$, $N_{rx}^0=0$
\FOR{i=1,i$\leqslant$ time slot amount,i++}
 \IF{$N_{rx}(i) \geqslant N_T(i)$}
    \STATE receive a "1" bit
    \STATE $N_{rx}^1 = N_{rx}^1 + N_{rx}(i)$, $n_1$++
 \ELSE
    \STATE receive a "0" bit
    \STATE $N_{rx}^0 = N_{rx}^0 + N_{rx}(i)$, $n_0$++
 \ENDIF

\STATE Calculate $A(i)=N_T(i)-\frac{N_{rx}^0}{n_0}$
\STATE Calculate $B(i)=\frac{N_{rx}^1}{n_1}-N_T(i)$

 \IF{$A(i)-B(i)>\mu$}
     \STATE $N_T(i+1)=N_0(i)-1$
 \ELSIF{$A(i)-B(i)<-\mu$}
     \STATE $N_T(i+1)=N_0(i)+1$
 \ELSE
     \STATE $N_T(i+1)=N_0(i)$
 \ENDIF

\ENDFOR
\end{algorithmic}
\end{algorithm}

A serious problem in MC is the communication reliability. Compared with traditional communication, biological channel condition of MC is more probabilistic, that may result in the high bit error rate (BER). The optimal threshold $N_T^{opt}$, that enables the BER lowest, is different under the various biological channel conditions. Moreover, $N_T^{opt}$ is also changing with the various bits because of the noisy randomness character. The receiver doesn't know how to set the best threshold at the beginning of the communication without knowledge of channel condition. So the half of the molecule emitted per bit $M/2$ is set as the threshold originally based on the typical experience, while it is not optimal that is proved in proposition 3. That's the motivation to design ATV algorithm, which aims to improve BER performance by adjusting the threshold. Adopting ATV, the receiver threshold changes naturally based on knowledge of received bits, which are related with channel condition. It is similar with the learning process, which makes receiver fit the realistic MC channel environment to demodulate properly.

The ATV is shown in the table Algorithm 1. The threshold of $i^{th}$ slot is denoted by $N_T(i)$. It varies dynamically, according to the distance between the previous mean received molecules and the previous thresholds. First of all, we set the initial threshold $N_T(1)$ as $M/2$, and two counters are set up, respectively counting the quantity of bit 1 and 0. Under the equivalent probability of emitting 1 and 0, the probability of receiving 0 and 1 is not equal, which will be shown in Figure 3. And the fixed threshold may result in the increase of the error probability, as the time lapses. Inspired by proposition 1, the optimal threshold should be the midpoint of the mean quantity of bit "1" and "0". But they change with time, so we calculate current the distance between the current threshold and the two mean values of bit "1""0", expressed as $A(i)$ and $B(i)$. If the distance difference of $A(i)$ and $B(i)$ exceeds the tolerant interval $\mu$, we treat that the balance of bit 1 an 0 is broken and the threshold will accordingly be regulated.

\subsection{ATV algorithm discussion}
Imaged that nano-devices range from nanometer to micrometer scale. Restrained by the tiny size and low power, not many resources are available to implement the sensing, communication, acting functions. So to fit the character, any algorithms are required to be simple, efficient. ATV algorithm contains three major steps. The first step is to set the counter of bit "1" and "0", and memory the molecule quantity per time slot, which ranges from line 1 to line 9. The time complexity of the first step is $n+1+nn_0+nn_1$, which equals $2n^2+n+1$. The second step is to calculate the numerical distance between current threshold and average molecules received for bit "1""0", that ranges from line 10 to 11. The time complexity of the second step is $2n^2$. The third step is the threshold variation process, ranging from line 12 to 19, of which the time complexity is $n^2$. Adding the three parts together, the entire complexity is $5n^2+n+1$. So We have the time complexity of ATV as,

\begin{equation}
T(n) = O(n^2)
\end{equation}

In ATV algorithm, only one "for loop" is used. Compared with some other existing methods which improve reliability in molecular communication, the best advantage of ATV is its simplification, low time complexity, that make it more likely to be utilized in nanoscale devices. We can infer that the varying threshold is convergent with a long enough time. The reason is that the signal and ISI components of $N_{rx}$ approach their fixed means when calculating $N_{rx}^1$ and $N_{rx}^0$. The random impact of the noise term to $A(i)$ and $B(i)$ becomes smaller and smaller with the increase of the $N_{rx}$. So $A(i)$ and $B(i)$ have a limit, that makes the variation of the threshold convergent.

ATV algorithm aims to keep the balance of demodulation, which are able to reduce the error probability of demodulate bit "1" or "0". Channel noise is an important factor to rise the BER. As we describe before, the noise is assumed as the AWGN for demodulation. Serious noisy channel here means a noise with a larger variance, which may be caused by crowded communication nano-devices. Mapping into the mathematical models of (14), the receiver is possible to absorb or lose more molecules. Hence the value of $A(i)-B(i)$ in ATV has a larger variable range, which makes $N_T(i)$ easier to change. Stronger noise also means a lager random factor, that lead the delay of threshold converge. Differently, in a good channel condition, there only exists slight noise. If the slot length is chosen properly, compared with signal, ISI is relatively weaker in terms of power. The threshold variation will not be fiercely and is easier to converge.

\section{PERFORMANCE EVOLUTION}
In this section, we simulate the reliability of free diffusion-based channel detailed in previous Sections. Considering the ISI and channel noise, BER of the channel is presented with some important parameters. The performance of the ATV algorithm is also verified. Note that we only consider the case that $n_t=n_r$. Let $\gamma_e$ be the signal to interference ratio (SINR), that is calculated based on the molecule ratio of the three branches in Section 4, expressed as,

\begin{equation}
\begin{split}
\gamma_e&=\frac{P_{signal}}{P_{isi}+P_{noise}}\\
&\approx \frac{\frac{1}{n}\sum_{n_t=1}^n|N(n_t,n_r)|^2}{\frac{1}{n}\sum_{n_t=1}^n|N(n_t+1,n_r)|^2+[\sigma^2]}
\end{split}
\end{equation}
Where $P_{signal}$, $P_{isi}$ and $P_{noise}$ respectively means the power of received signal, ISI signal and channel noise. They are established like this because molecule quantity is a discrete value. [.] is the floor operator. (15) reflects the ratio of each part of molecules directly. When receiver counts molecules in every time slot, ISI and channel noise molecules result in the quantity inaccuracy and promote error bit determine. And they could be considered as the addictive signals comparing with receiving signals according to the character of concentration modulation and quantity threshold decision. $n$ is the total quantity of the time slot, and $P_{noise}$ is its variance since it follows a zero-mean Normal distribution.

\begin{table}
\centering
\caption{SIMULATION PARAMETERS}
\begin{tabular}{|c|c|l|c|} \hline
Diffusive Coefficient&$D$&10,1000&$\mu m^2/s$\\ \hline
Transceiver distance&$r$&1-20&$\mu$\\ \hline
Time slot length&$\tau$&1-10&$s$\\ \hline
Binding rate of the receptors&$a$&0.1&/\\ \hline
Releasing rate of the receptors&$b$&0.08&/\\ \hline
Concentration of the receptors&$Q$&1&$umol/l$\\ \hline
Power of AWGN&$P_{noise}$&1-20&$\mu W$\\ \hline
Receiver threshold&$N_T$&50-450&/ \\ \hline
Molecules for bit "1"&$M$&500&/ \\ \hline
Tolerant interval&$\mu$&30,60&/ \\
\hline\end{tabular}
\end{table}

\begin{figure}
\centering
\noindent\makebox[\textwidth][c] {
\includegraphics[width=7in]{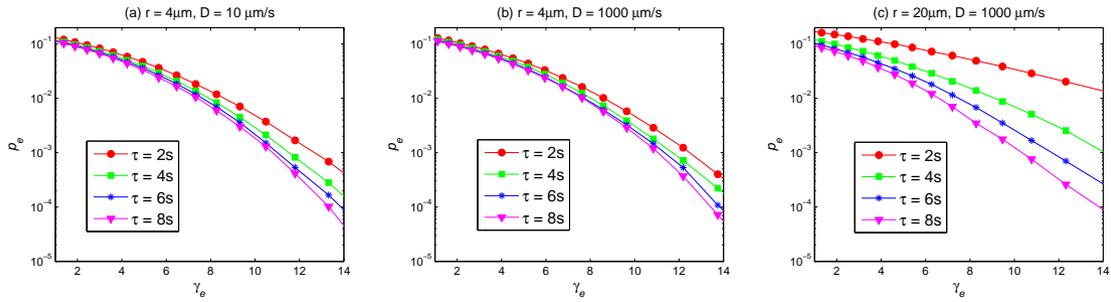}}
\caption{BER of communication for different $D$, $r$ and $\tau$ under various $\gamma_e$}
\end{figure}

\subsection{Simulation Design}
The major simulation parameters are listed in table 1. These parameters could be divided into two branches, i.e., diffusion-based channel parameters and signal (including interference and noise) parameters. In [31], the diffusion coefficient $D$ is considered as the range of 10-1000 $\mu m^2/s$. The channel conditions under different $D$ are verified. In this paper, we study the diffusive channel based on some discrete $D$. We set the transceiver distance $r$ as 1-20 $\mu m$, similar with the references [29] and [31]. The value of $a$,$b$ and $Q$ indicate the parameter of the ligand-based model, that are derived from [21]. In our paper, molecules quantity per bit $M$ is set as 500. The setup of $\mu$ mainly refers to the result of Figure 5 and 6. We set it as 30 and 60, when the time slot length is 2s.

\subsection{Numerical analysis}
\begin{figure}
\centering
\includegraphics[width=3.5in]{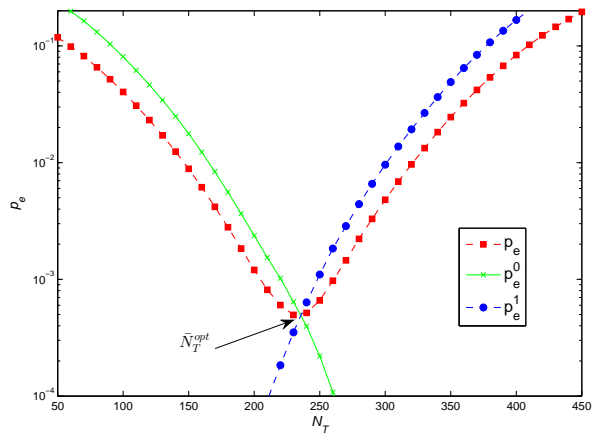}
\caption{BER and Error rate of bit "1""0", $\tau$=4s, $D$=10$\mu m^2/s$, $r$=4$\mu m$, $\gamma_e$=10}
\end{figure}

\begin{figure}
\centering
\includegraphics[width=3.5in]{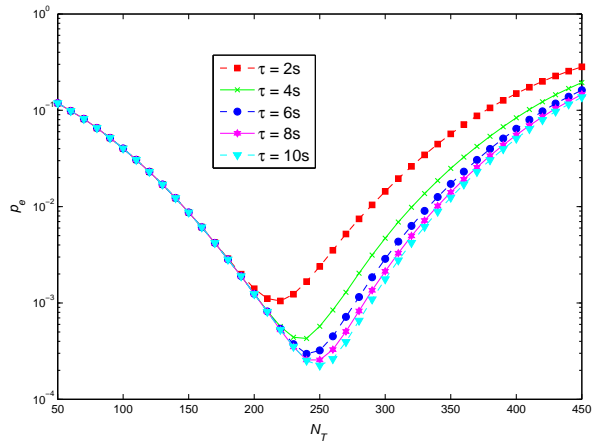}
\caption{BER for various $N_T$ and $\tau$, $D$=10$\mu m^2/s$, $r$=4$\mu m$, $\gamma_e$=10}
\end{figure}

Figure 3 illustrates the relationship between BER $p_e$ and channel diffusion $D$, transmission distance $r$, as well as time slot length $\tau$. The $\gamma_e$ varies through altering the noise power. We can see that $p_e$ decrease with $\gamma_e$ increase. There is little difference for $p_e$ of different $\tau$ when $\gamma_e$ is low. Comparing (a) and (b), we could get the conclusion that the BER performance does not have a significant difference for different $D$. While comparing (b) and (c), larger the distance $r$ is, worse the BER performance is. This attributes to two factors. Firstly, increase of $r$ directly results in decline of signal molecule arriving probability, which refers to (3). Correspondingly, the impact to ISI molecule is no far more significant than signal molecules. Moreover, for a longer $r$, the signal molecules decrease obviously, so channel noise will impact more when receiver counts molecule per time slot.

\begin{figure}
\centering
\includegraphics[width=3.5in]{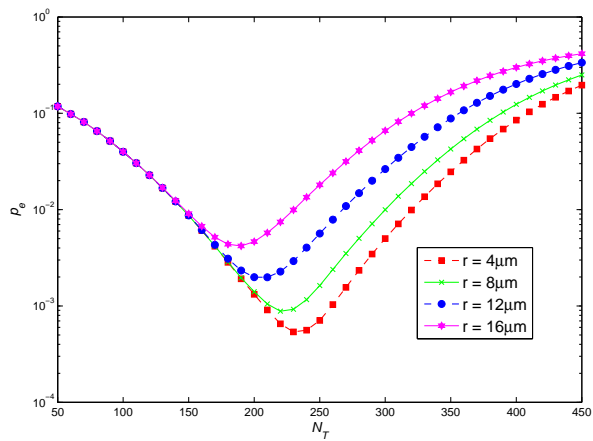}
\caption{BER for various $N_T$ and $r$, $D$=10$\mu m^2/s$, $\tau$=4$s$, $\gamma_e$=10}
\end{figure}

Based on Figure 4, we can see that $p_e^0$ and $p_e^1$ decrease and increase respectively with the increase of $N_T$. The crossing point of the two curves decides the optimal value of $N_0$. It is smaller than the midpoint, i.e., $N_{tx}/2$ according to the simulation. If current $N_0$ is much greater than the optimal value, $p_e^0$ falls lightly while $p_e^1$ rises heavily, as a result the BER becomes large. We can get the similar conclusion if $N_T$ is much less than the optimal value.

In Figure 5, the BER $p_e$ is shown with various $N_T$ under different time slot $\tau$. $p_e$ first decreases and then increases, with enlargement of the threshold. For different $T$, the optimal threshold $N_T^{opt}$ is different. The molecule quantity per bit $M$ is set as 500, while $N_T^{opt}$ is lower than half of the molecule amount per bit, i.e. 250, owing to the specificity of molecular OOK modulation. It also presents that $N_T^{opt}$ trends to get a greater value with the increase of $\tau$. Correspondingly, Figure 6 reveals the relationship between BER and threshold under different $r$. The variation trend for threshold is the same with Figure 5. However, $p_e$ is lower for lager $r$, and $N_T^{opt}$ is also farther from the middle point. Hence, we can conclude that for various $\tau$ and $r$, the threshold for minimum $p_e$ is various.

\begin{figure}
\centering
\includegraphics[width=3.5in]{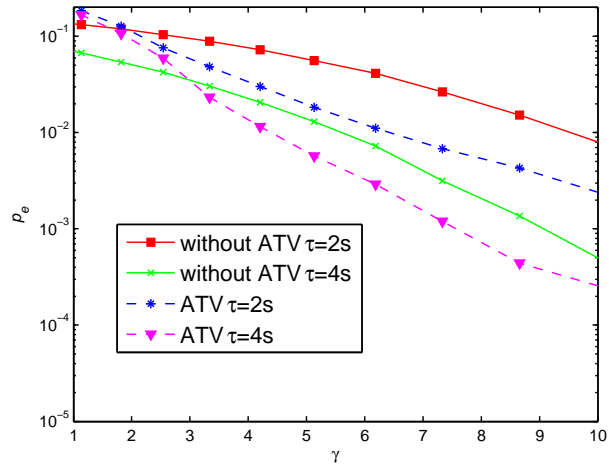}
\caption{Comparison of using or not using ATV under various $\gamma_e$, $D$=10$\mu m/s$, $r$=8$\mu m$}
\end{figure}

In Figure 7, the BER performance of using or not algorithm 1 under various $\tau$ is demonstrated. Although when $\gamma_e$ is extremely low, the performance improvement is not obvious, even worse. However, when $\gamma_e$ is higher, use of ATV algorithm could lead a distinct difference. The receiver could not obtain the right amount of the signal molecules and the proper threshold will be deviated to the original value. The ATV algorithm makes the actual threshold closer to theoretic optimal threshold, so the $p_e$ decrease. In addition, for a shorter $\tau$, adopting ATV could get a more significant BER performance, gaining smaller channel delay.

\begin{figure}
\centering
\includegraphics[width=3.5in]{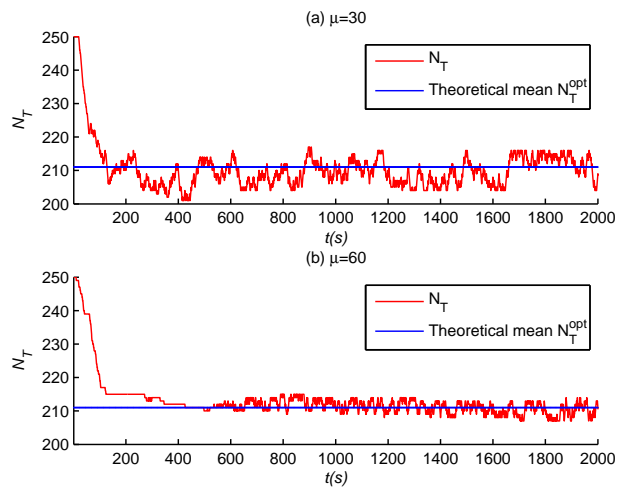}
\caption{Receiver threshold varied with time, $\tau$=2s, $D$=10$\mu m^2$/s, $\gamma_e$ = 10}
\end{figure}

Figure 8 and 9 show the variation of receiver threshold if the ATV algorithm is adopted. In those figures, we can see that the threshold decreases sharply firstly. There exists much sawteeth due to the randomness of AWGN and molecular Brownian motion. Comparing figure 6 (a) and (b), the first curve shakes a little heavier, just because that the tolerant interval $\mu$ is smaller, leading threshold changing easier. For Figure 8 (a) and Figure 9 (c), only the condition of $\gamma_e$ is different. We could see that, lower $\gamma_e$, i.e., worse communication environment, causes much more shake of the varied threshold. So the $p_e$ will also increase accordingly. We could conclude that in both good and a little worse communication condition, the threshold of ATV is more closer to best theoretic threshold.

\begin{figure}
\centering
\includegraphics[width=3.5in]{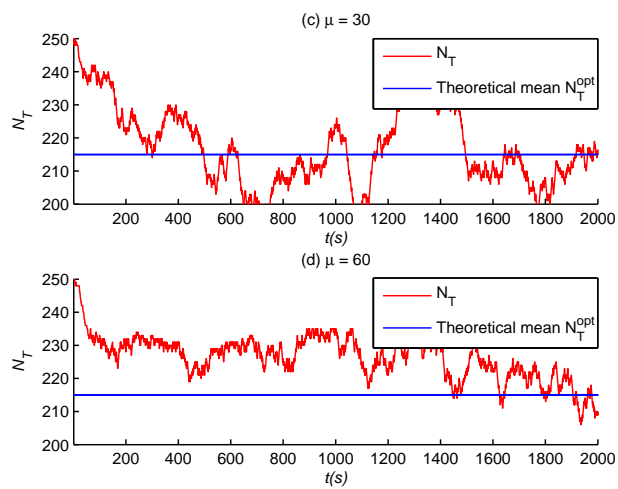}
\caption{Receiver threshold varied with time, $\tau$=2s, $D$=10$\mu m^2$/s, $\gamma_e$ = 5}
\end{figure}

\section{CONCLUSIONS}
In this paper, we investigate the reliability of the diffusion-based molecular channel. We first describe the basic system model. A classified model is established to divide molecules into signal, ISI and noise parts in demodulation. We expand the three parts respectively and connect them using a traditional-like formula. Based on the classified model, we do a theoretical analysis, which give the basis for the later ATV algorithm design. The ATV algorithm at the receiver is designed to improve the BER performance. We discuss its complexity and performance on various noisy channels. Then, we verify our proposal through simulation. The results show that the algorithm decreases BER obviously as long as the channel condition is not too bad. The threshold of the receiver is more closer to the best theoretic threshold, so the algorithm works. This work contributes to improve the communication reliability for diffusion-based molecular communication.

\end{document}